\documentclass[manuscript]{emulateapj}

\usepackage{natbib}
\bibpunct{(}{)}{,}{a}{}{,}

\slugcomment{Ver 2, 31 Jan 2016}

\shorttitle{Heat as an inertial force}
\shortauthors{Thanjavur \& Israel}

\begin{document}
\def\asec{^{\prime\prime}}
\def\mag{\:{\mathrm mag}}
\def\angs{\mathrm{\AA}}
\def\sqdeg{\:\mathrm{deg^2}}
\def\deg{\ensuremath{^\circ}}
\def\omegam{$\Omega_{\mathrm m}$}
\def\omegab{$\Omega_{\mathrm b}$}
\def\omegabh{$\Omega_{\mathrm b} h^2$}
\def\omegal{$\Omega_{\Lambda}$}
\def\ovrdn{$\Delta_{200}$}
\def\ovrdnr{$\Delta(r)$}
\def\ovrdnv{$\Delta_{vir}$}
\def\vcr{$v_{\mathrm c}(r)$}
\def\vc{$v_{\mathrm c}(r)$}
\def\halpha{$\mathrm H_{\alpha}$}
\def\rhoc{$\rho_{\mathrm crit}$}
\def\rhocore{$\rho_{0}$}
\def\rcore{$r_{\mathrm c}$}
\def\rhobar{$\bar{\rho}$}
\def\rhor{$\rho(r)$}
\def\mr{$M(r)$}
\def\radius{$r$}
\def\rscale{$r_{\mathrm s}$}
\def\deltac{$\delta_{\mathrm c}$}
\def\lcdm{$\Lambda$CDM }
\def\scdm{$S$CDM }
\def\rnfw{$r^{-1}$}
\def\mathrmoore{$r^{-1.5}$}
\def\hub{$h$}
\newcommand{\mc}[3]{\multicolumn{#1}{#2}{#3}}
\def\G{{\mathrm G}}
\def\c{{\mathrm c}}
\def\gsim{ \lower .75ex \hbox{$\sim$} \llap{\raise .27ex \hbox{$>$}} }
\def\lsim{ \lower .75ex \hbox{$\sim$} \llap{\raise .27ex \hbox{$<$}} }
\def\Mpc{{\mathrm Mpc}}
\def\kpc{{\mathrm kpc}}
\def\pc{{\mathrm pc}}
\def\Lsun{{\mathrm L_\odot}}
\def\Msun{\:{\rm M_\odot}}
\def\gsim{\ga}
\def\eg{e.g.\ }
\def\lsim{\la}
\def\etal{{et al.\ }}
\def\mpc {\mathrm{Mpc}}
\def\kpc {\mathrm{kpc}}
\def\msun {\:{\rm M_\odot}}
\def\ergs {{\mathrm erg} \, {\mathrm s}^{-1}}
\def\cm {{\mathrm cm}}
\def\kms {{\mathrm {\,km \, s^{-1}}}}
\def\Hz {{\mathrm Hz}}
\def\yr {{\mathrm yr}}
\def\gyr {{\mathrm Gyr}}
\def\arcmin {{\mathrm arcmin}}
\def\G {{ G  }}
\def\magasec {\mathrm {\, mag.arc.sec^{-2}}} 
\def\lsim{\mathrel{\hbox{\rlap{\hbox{\lower4pt\hbox{$\sim$}}}\hbox{$<$}}}}
\def\gsim{\mathrel{\hbox{\rlap{\hbox{\lower4pt\hbox{$\sim$}}}\hbox{$>$}}}}

\def\and   {\mathrm {et al.} \mathrm}  
\def\mathrmd {\mathrm d}



\title{HEAT AS AN INERTIAL FORCE:  A QUANTUM EQUIVALENCE PRINCIPLE}

\author{Karun Thanjavur\altaffilmark{1} and Werner Israel\altaffilmark{1}}

\altaffiltext{1}{Department of Physics \& Astronomy, University of Victoria, 
        Victoria, BC, V8P 1A1, Canada; karun@uvic.ca, israel@uvic.ca}

\begin{abstract}
The firewall was introduced into black hole evaporation scenarios as a \textit{deus ex machina} designed to break entanglements and preserve unitarity \citep{AMPS13}. Here we show that the firewall actually exists and does break entanglements, but only in the context of a virtual reality for observers stationed near the horizon, who are following the long-term evolution of the hole. These observers are heated by acceleration radiation at the Unruh temperature and see pair creation at the horizon as a high-energy phenomenon. The objective reality is very different. We argue that Hawking pair creation is entirely a low-energy process in which entanglements never arise. The Hawking particles materialize as low-energy excitations with typical wavelength considerably larger than the black hole radius. They thus emerge into a very non-uniform environment inimical to entanglement-formation.

\end{abstract}



\section{Introduction \label{intro}}

\noindent A man whirling around on a carousel finds it convenient to talk about a centrifugal or inertial force. The force is not real, but it provides an internally consistent account of dynamics in his frame if he chooses to ignore the fact that he is accelerating. Einstein's elevator is a well-known example of the fertility of such concepts in classical physics.

\indent When quantum effects come into play, matters become more interesting. Forty years ago, Unruh showed that a particle detector accelerating in vacuum reacts just as if it were at rest but in contact with a thermal bath at a temperature proportional to its acceleration \citep{Unru76}. The bath is fictitious, but it provides an internally consistent description of physics in the detector's rest-frame if we choose to ignore its acceleration.

\indent All this becomes relevant to Hawking evaporation when we recall that our conception of pair creation by black holes and the resulting entanglements leans quite heavily on a simple mental picture. Two entangled excitations of very high energy emerge at or near the horizon. One falls into the hole; the other escapes and gets red-shifted to become a Hawking particle with typical energy on the order of the Hawking temperature.

\indent But the formal theory lends no direct support to this simple picture. Hawking's original derivation \citep{Hawk75} was based on an asymptotic S-matrix approach. Evaluations of the stress-energy tensor in the Unruh state show that expectation values are everywhere finite, with no sign of high energy at the horizon \citep{Fabb05}.

\indent Seemingly at odds with this is the fact that a particle detector stationed near the horizon becomes highly excited \citep{Fabb05}.  But this detector is statically supported against gravity and therefore strongly accelerated. Its response is due entirely to the Unruh effect \citep{Unru76}. We have no reason to think that any of it stems from the black hole, and have no evidence for a high-energy origin of the Hawking particles.

\indent These intrusive heating effects of acceleration are the price one pays if one is studying physics near the horizon from a stationary platform. But they are unavoidable if we want to follow black hole evaporation over a long stretch of time. They are inertial effects in the same sense as the inertial forces of classical mechanics, and must be taken into account for an internally consistent description  of the static observer's experience. They are an inseparable part of his \textit{virtual} reality.

\indent Hawking excitations emerging from the horizon in this virtual narrative are at once enveloped in a bath of acceleration radiation (the \textit{firewall}, located on a sphere at the radius of the observer) and heated to relativistic energies. They become in fact the excitations of our simple mental picture. And the firewall will abort any interactions that might otherwise arise between outgoing partners of early and late pair-creation events. (We recall that this was the crux of the AMPS bigamous entanglement paradox (\citet{AMPS13}, henceforth \textit{AMPS13}).

\indent This narrative, although quite divorced from reality, is internally self-consistent, and correctly predicts observable effects -- in particular, that bigamous entanglements never occur.

\indent But clearly this is not the way it actually happens. The firewall is only a fiction, an artefact of the static observer's virtual reality.

\indent In the real, low-energy scenario we are advocating here, the partners of a pair-creation event materialize with wavelengths substantially larger than the size of the hole, i.e., into  very nonuniform surroundings. Entanglements will easily form only if the environment allows easy exchange of the partners' positions, which is not the case here. It seems plausible that no entanglements will arise in the first place.
 
\indent Thus the $AMPS13$ bigamous entanglement paradox is resolved in both the real and virtual scenarios, though in completely different ways. However, it must be emphasized that the virtual scenario may be fictitious, but it is very far from being superfluous and useless. Because it is an elegantly clean and consistent scenario, its predictive power is very much greater. For instance, it makes the correct observable prediction that the Hawking spectrum is thermal. This is completely lost in the fuzzy picture of the real scenario seen by a free-falling observer. Therefore we go as far as to claim that the quantum equivalence principle (QEP) is not just a magnificent cultural ornament, but is directly applicable and useful to science.

\indent Our argument can be strengthened to provide a proof that information cannot be lost in black hole evaporation. In the static observer's virtual reality, information falling toward the hole is intercepted by the Unruh firewall and reradiated to the outside. None of it enters the hole in this scenario, and none is lost. QEP then guarantees that information is ultimately preserved also in the real scenario, though in a much more complicated way.

\indent QEP bears a superficial resemblance to BH Complementarity \citep{Suss08}, long championed by Leonard Susskind. Both contrast the experiences of static and free-falling observers near the BH. However, QEP asserts that the static experience is not complementary in the sense of Bohr, but completely different and fictitious (``virtual reality"). And it exposes and emphasizes the key role of Unruh heating in the contrast. But in their operational philosophy, the two approaches have much in common.

\section*{Summary \label{sum}}
\noindent  It is worth restating in plain terms our viewpoints set forth above:
\begin{enumerate}
\item In the real world, there is no firewall at the horizon. A real firewall would have unacceptable consequences. 
\item Pair creation by a black hole is not a localized process confined to the horizon; it spreads over a broad region that includes the horizon.
\end{enumerate}

\noindent Hawking excitations, emerging from the horizon in the stationary observer's virtual narrative, are at once enveloped in acceleration radiation and heated to relativistic energies. They become, in fact, the excitations of  our simple mental picture. To a network of  stationary observers distributed over a sphere of radius r, this will appear as a \textit{firewall} of radius r. And the firewall will abort any interactions that might otherwise arise between outgoing partners of early and late pair-creation events. This was the crux of the \textit{AMPS13} bigamous entanglement paradox (\citet{AMPS13}.
                        
\indent Bardeen has argued forcefully that Hawking radiation must have a low-energy, non-local origin \citep{Bard14}. And it does seem a bit odd that high-energy particles should pop out in low-curvature vacuum surroundings from a surface that is only defined globally. The snag for this conservative but unfashionable view has always been: how is it, then, that the radiation emerges near infinity with a thermal spectrum stamped with the signature of the horizon (its surface gravity), in perfect agreement with Hawking's asymptotic derivation \citep{Hawk75}?  And why, then, should the Kraus-Wilczek amplitudes \citep{Krau94} for tunnelling through the horizon match the thermal results? 

\indent The ``virtual reality'' scenario provides an answer to both questions. 

\indent This paper has no formulas, and for a reason. The formal theory, as far as it goes, is in good shape; we do not tamper with it. The problem is that the formal theory is still incomplete, and we have been bridging the gaps with an intuitive picture. And the intuitive picture has been misleading us because it, too, is incomplete.        

\indent When we talk about entanglement formation and disruption, this is not, or at least not yet, based on formal calculation. It is plausible guesswork. We are suggesting that the intuition behind this guesswork is faulty, incomplete, because it leaves out a key ingredient, the heating effects of acceleration.            

\indent Take an operational view: we sit on a platform stationed near the black hole and watch for a long time as it evaporates. But then we get a distorted picture of what is happening to the black hole itself, because our platform is accelerating, and this has consequences, especially quantum consequences.
 
\indent Unruh taught us that the quantum consequences can be thought of as a heating effect \citep{Unru76}. For us, this heating is just as ``real'' as centrifugal force is for someone on a carousel. For him, it ``explains'' why loose objects on the carousel fly outwards, and for us, it ``explains'' the subtle quantum effects of our acceleration on the things we perceive and the conclusions we draw from those perceptions. What we perceive is an outflow of ``hot'' particles (Unruh-heated Hawking excitations) issuing from a neighbourhood of the horizon---very much our traditional picture of pair creation at the horizon---but now laced with a new feature: Unruh heating, a ``firewall''.

\indent Objectively, this heating effect does not exist. It makes no contribution to the material stress-energy and it has no effect on the space-time geometry. But, within the context of a consistent ``virtual narrative'' spun by a static observer near the horizon who chooses to ignore effects of his acceleration, it \textit{does} exist. And it has the power to act on the other elements of his virtual narrative. In particular, it can abort or break virtual entanglements formed in his narrative of pair creation.

\indent In this view of things, the firewall appears no longer as an ad hoc intervention, but as a built-in entanglement-slasher. Bigamy is not an option.

\indent In short: our conventional picture of high-energy pair creation at the horizon is a fiction. But we can adopt this picture and still predict observable phenomena correctly, provided we accept the baggage that comes with it: a firewall (in reality, created by our acceleration).

\indent To conclude on a speculative note: Einstein's equivalence principle served as a bridge between Newtonian gravity and classical general relativity. Could a thermally extended quantum version, of the sort adumbrated here, play a similar role as passport to a successful quantum theory of gravity? Perhaps. Optimism is tempered by Einstein's admonition: ``A good joke should not be repeated too often''. But that was addressed to Heisenberg, and concerned his quantum uncertainty principle.

\section*{Acknowledgement}         

\indent Our warm thanks to Jim Bardeen for his incisive, illuminating exchange which helped us improve our arguments in this revised manuscript. One of us (W.I.) has enjoyed the support of the Canadian Institute for Advanced Research for thirty years. He is indebted to colleagues in the CIFAR Cosmology and Gravity Program, in particular Bill Unruh  and Frans Pretorius, for many stimulating interactions.





\bibliographystyle{apj}
\bibliography{Firewallv2}

\clearpage

\end{document}